\documentclass{article}

\usepackage{arxiv}

\usepackage[utf8]{inputenc} % allow utf-8 input
\usepackage[T1]{fontenc}    % use 8-bit T1 fonts
\usepackage{url}            % simple URL typesetting
\usepackage{booktabs}       % professional-quality tables
\usepackage{nicefrac}       % compact symbols for 1/2, etc.
\usepackage{microtype}      % microtypography
\usepackage{lipsum}
\usepackage{graphicx}
\graphicspath{ {./images/} }

\usepackage[colorlinks=true, allcolors=blue]{hyperref}
\usepackage{amsmath,amssymb}
\usepackage{graphicx}
\usepackage{wrapfig}
\usepackage{multirow}
\usepackage{enumitem}
\usepackage{caption, subcaption}

\title{Deep MRI Reconstruction with Radial Subsampling}

\author{
 George Yiasemis \\
  Netherlands Cancer Institute,\\
  Amsterdam, 1066 CX,\\
  the Netherlands \\
  \texttt{g.yiasemis@nki.nl} \\
   \And
      Chaoping Zhang \\
   Netherlands Cancer Institute,\\
  Amsterdam, 1066 CX,\\
  the Netherlands \\
  \texttt{c.zhang@nki.nl} \\
   \And
 Clara I. Sánchez \\
  qurAI group\\
  University of Amsterdam,\\
  Amsterdam, 1012 WX, \\
  the Netherlands \\
  \texttt{c.i.sanchezgutierrez@uva.nl} \\
  \And
 Jan-Jakob Sonke \\
  Netherlands Cancer Institute,\\
  Amsterdam, 1066 CX,\\
  the Netherlands \\
  \texttt{j.sonke@nki.nl} \\
   \And
   Jonas Teuwen \\
   Netherlands Cancer Institute,\\
  Amsterdam, 1066 CX,\\
  the Netherlands \\
  \texttt{j.teuwen@nki.nl} \\
}

\begin{document}
\maketitle
\begin{abstract}
In spite of its extensive adaptation in almost every medical diagnostic and examinatorial application, Magnetic Resonance Imaging (MRI) is still a slow imaging modality which limits its use for dynamic imaging. In recent years, Parallel Imaging (PI) and Compressed Sensing (CS) have been utilised to accelerate the MRI acquisition. In clinical settings, subsampling the $k$-space measurements during scanning time using Cartesian trajectories, such as rectilinear sampling, is currently the most conventional CS approach applied which, however, is prone to producing aliased reconstructions. With the advent of the involvement of Deep Learning (DL) in accelerating the MRI, reconstructing faithful images from subsampled data became increasingly promising. Retrospectively applying a subsampling mask onto the $k$-space data is a way of simulating the accelerated acquisition of $k$-space data in real clinical setting. In this paper we compare and provide a review for the effect of applying either rectilinear or radial retrospective subsampling on the quality of the reconstructions outputted by trained deep neural networks. With the same choice of hyper-parameters, we train and evaluate two distinct Recurrent Inference Machines (RIMs), one for each type of subsampling. The qualitative and quantitative results of our experiments indicate that the model trained on data with radial subsampling attains higher performance and learns to estimate reconstructions with higher fidelity paving the way for other DL approaches to involve radial subsampling.
\end{abstract}


\section{Introduction}
\label{sec:intro}
Magnetic Resonance Imaging (MRI) is the most widely used medical imaging modality employed to produce high quality detailed images of both the anatomy and physiology of the human body. The fact that it allows for non-invasive clinical examinations and diagnosis, has played a significant role in its widespread use. MRI however, suffers from very long acquisition times as a result of the fact that MRI data ($k$-space or Fourier domain) are acquired sequentially in time, frequently causing patient distress.  Additionally, these long acquisition times make it nearly intractable to deliver dynamic treatment such as image-guided targeted radiotherapy in real time. 

Truncating the MR imaging time, referred to as accelerating MRI \cite{zbontar2019fastmri}, is pivotal not only for improving patient comfort, decreasing scanning time and reducing medical costs, but, it can also make image-guided treatment feasible. Over the past fifteen years accelerating the MRI  acquisition has been in the center of many studies yielding advancements such as Parallel Imaging (PI) and Compressed Sensing (CS). 

\begin{figure}[ht!]
     \centering
      \begin{subfigure}[b]{0.24\textwidth}
         \centering
         \includegraphics[width=0.9\linewidth]{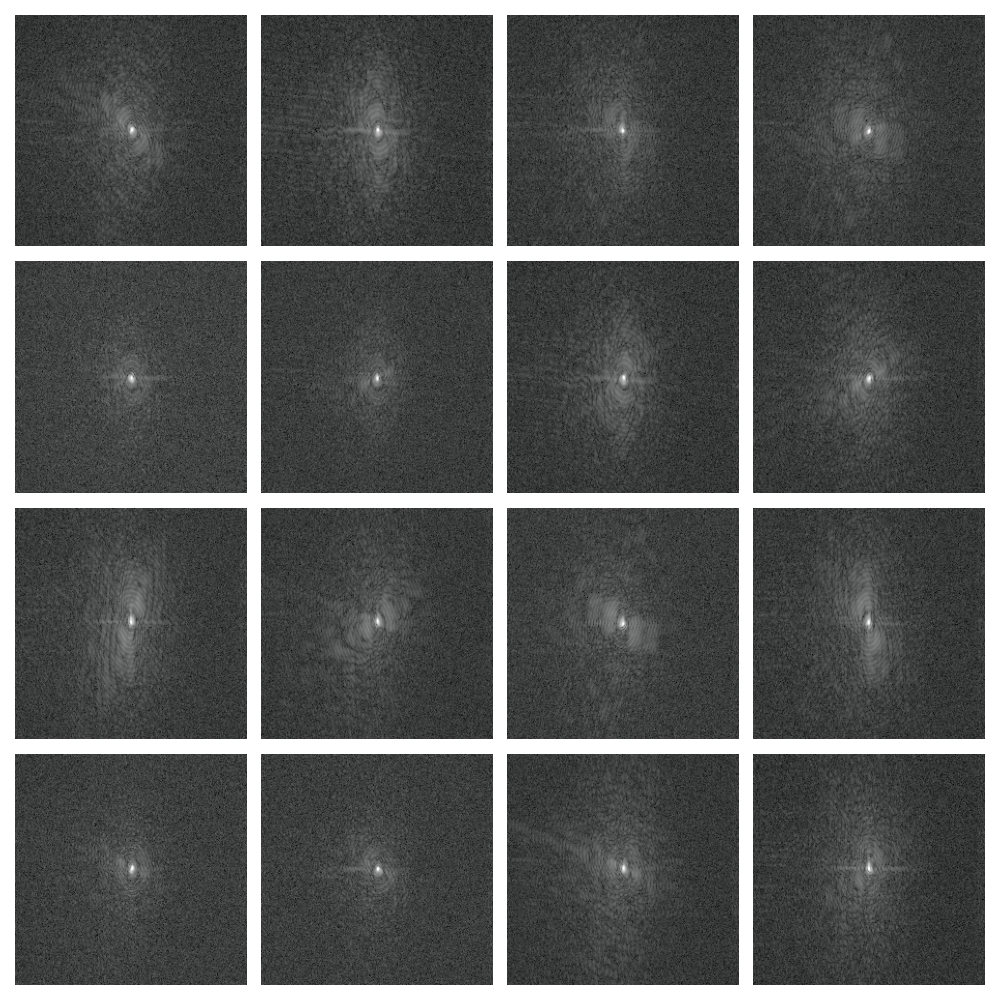}
         \caption{$k$-space data from sixteen receiver coils}
         \label{fig:kspace_coils}
     \end{subfigure}
     \hfill
     \begin{subfigure}[b]{0.24\textwidth}
         \centering
         \includegraphics[width=0.9\linewidth]{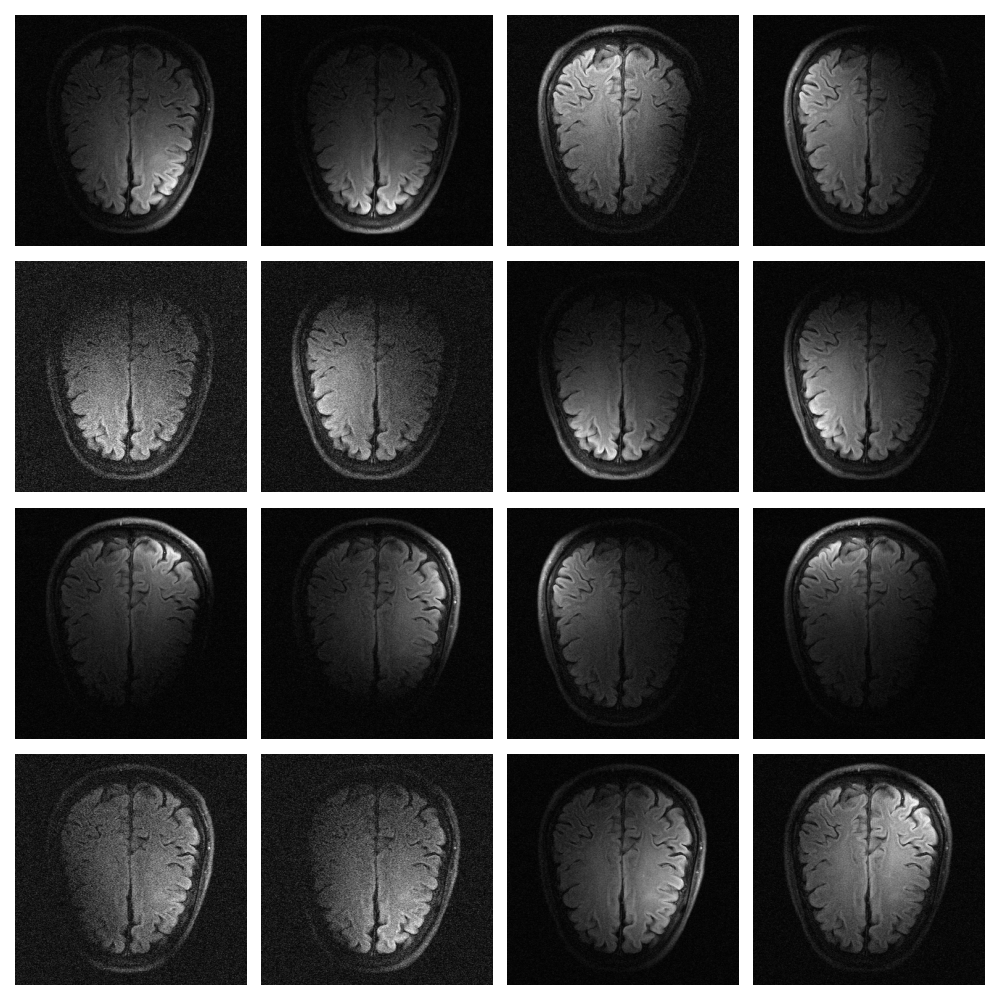}
         \caption{Coil spatial reconstructed images}
         \label{fig:coils}
     \end{subfigure}
     \hfill
      \begin{subfigure}[b]{0.24\textwidth}
         \centering
         \includegraphics[width=0.9\linewidth]{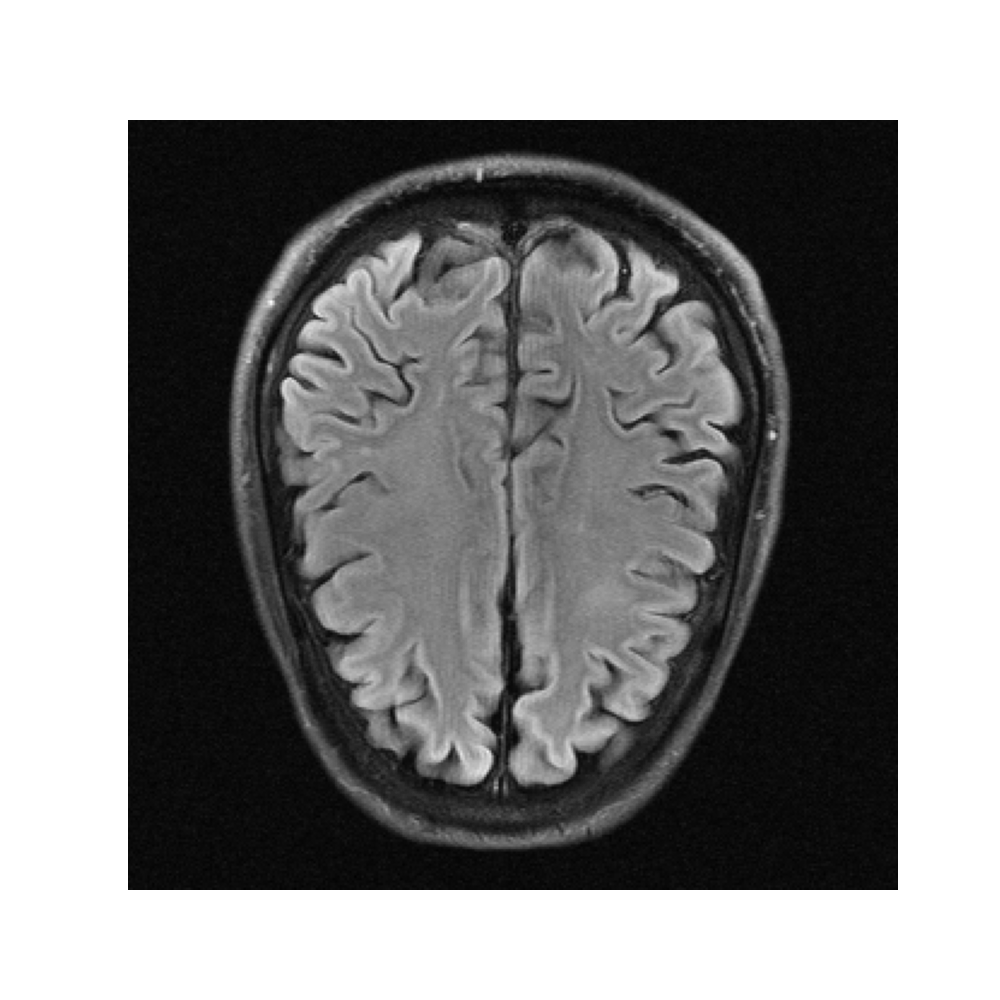}
         \caption{RSS reconstructed image from coil data}
         \label{fig:rec_from_coils}
     \end{subfigure}
     \hfill
     \begin{subfigure}[b]{0.24\textwidth}
         \centering
         \includegraphics[width=0.9\linewidth]{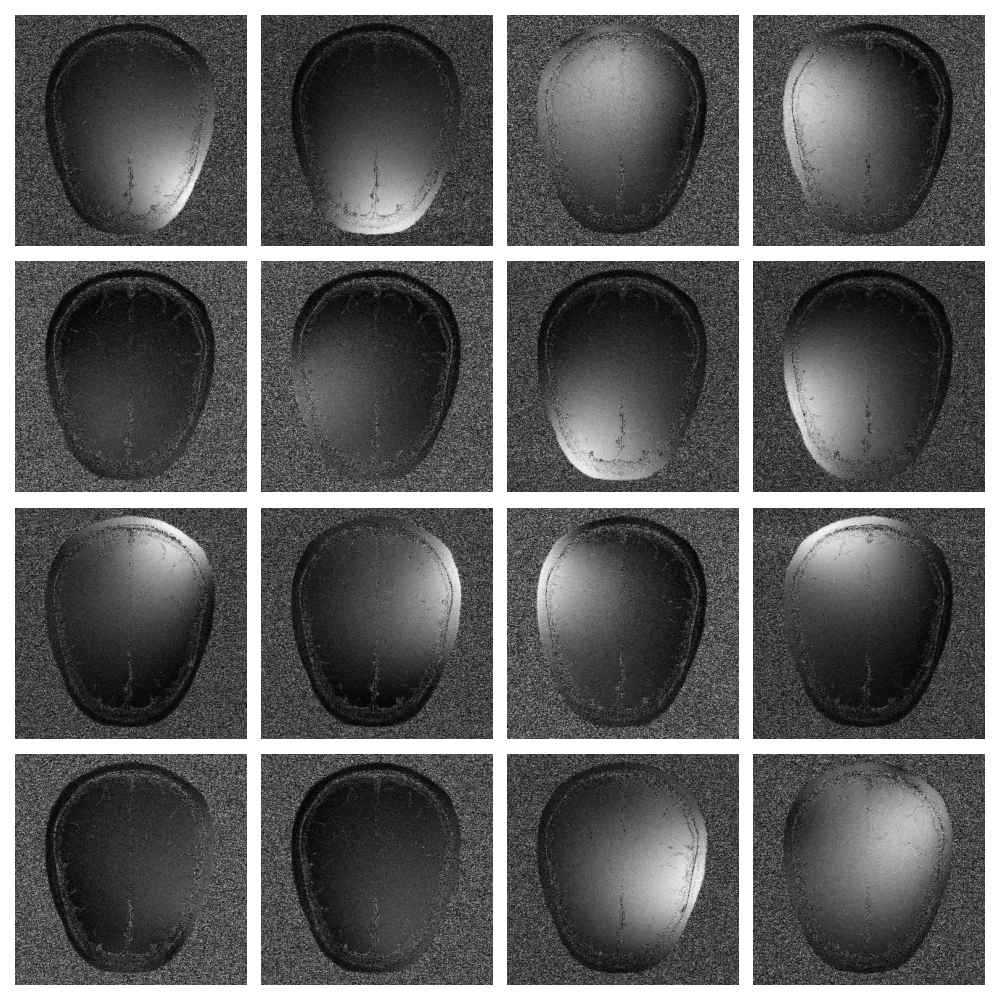}
         \caption{Coil sensitivity map magnitudes}
         \label{fig:sens_coils}
     \end{subfigure}
     \hfill
    \vspace{2.0mm}
    \caption{Parallel Imaging - MRI reconstruction using fully-sampled $k$-spaces from multiple coils.}
    \label{fig:coils&recon}
\end{figure}

In Parallel Imaging multiple radio-frequency receiver coils are utilised to read-out partial $k$-space measurements that depend on the position of each coil  \cite{parallel-https://doi.org/10.1002/nbm.1042}. A unique sensitivity map corresponds to each individual receiver coil and can be estimated with different techniques such as SENSE \cite{https://doi.org/10.1002/cmr.a.21244, senseencoding}. An example of a reconstructed brain image with the root-sum-of-squares (RSS) method acquired from multiple coils is illustrated in Figure \ref{fig:coils&recon}. 

Compressed Sensing is a technique used in clinical settings that aims in accelerating MRI acquisition by  acquiring sparse or partially observed $k$-space measurements  \cite{cs-1580791, cs-https://doi.org/10.1002/mrm.21391, cs-lustig-2, cs-https://doi.org/10.1002/cpa.20124, https://doi.org/10.1002/mrm.22428, 1614066}. In CS an acceleration factor $R$ is defined ($R$-fold acceleration) as the ratio of the number of the fully sampled $k$-space measurements to the number of measurements of the partially sampled $k$-space. Reconstructing partially observed $k$-space data violates the Nyquist sampling criterion \cite{doi:https://doi.org/10.1002/9781118633953.ch12}, therefore leading to aliased and/or blurred reconstructed images, also known as aliasing artifacts which in many cases can be identified but in others can be interpreted as pathology. 

In recent years, various Deep Learning approaches have been deployed aiming in reconstructing partially observed MRI data achieving state-of-the-art performance \cite{Aggarwal_2019, hammernik2017learning, zhang2019reducing, Hyun_2018, 8425639, yu2017deep, sriram2020grappanet, schlemper2017deep, han2019kspace}. The majority of these approaches train DL models on retrospectively subsampled measurements by applying rectilinear subsampling schemes (discussed further in Section \ref{sec:ch2}) which even with the use of DL produce imperfect reconstructions. 

In this work we investigate the performance of accelerated image reconstruction of a particular DL architecture (RIM) on retrospectively subsampled $k$-space data via two subsampling schemes, rectilinear and radial (Section \ref{par:subsampling}). We argue and show that the model trained on data subsampled with the latter scheme, can produce more precise and faithful reconstructions. In summary, the contributions of this paper are the following:
\begin{itemize}
    \item We are the first to employ the RIM architecture for training with radially subsampled data.
    \item We show through evaluation that the model we train utilising radial subsampling produces reconstructions with higher fidelity than the model trained with conventional rectilinear subsampling.
    \item We pave the way for other DL approaches to involve prospective or retrospective radial subsampling.
\end{itemize}

In Section \ref{sec:ch2}, we provide a mathematical formulation of the accelerated MRI reconstruction along with some information about prospective and retrospective subsampling. Section \ref{sec:ch3} focuses on our experiments implementation and demonstrates some of our results. In Section \ref{sec:ch4} we conclude and discuss our results and contributions. Our code is available at \url{https://github.com/NKI-AI/direct}.
% \newpage
\section{Background}
\label{sec:ch2}
\subsection{Mathematical Formulation}

In this section we briefly define the problem of reconstructing accelerated MRI data from multiple coils mathematically. 
Let $\mathbf{x}\,\in\,\mathbf{C}^{N \times M}$ denote the true image. For a number of $n_c$ coils, let $\mathbf{y}^k,\,\mathbf{S}^k\in\mathbb{C}^{N \times M}$ denote the fully acquired $k$-space measurements and sensitivity maps from coil $k$. Then,
\begin{equation}
    \mathbf{y}^{k}\,=\,\mathcal{F}(\mathbf{S}^{k} \circ \mathbf{x}) \,+\, \mathbf{e}^k,\quad k\,=\,1,\,..,\,n_c,
\end{equation}
where $\mathcal{F}$ denotes the discrete Fast Fourier Transform (FFT), $\mathbf{e}^k\in\mathbb{C}^{N\times M}$ the accumulated noise from coil $k$, and $\circ$ the element-wise product. In PI-CS, the RSS reconstruction is commonly used to reconstruct an image from the multiple $k$-space measurements \cite{https://doi.org/10.1002/mrm.1910160203} given by
\begin{equation}
    \Tilde{\mathbf{x}}\,=\,\text{RSS}(\Tilde{\mathbf{x}}^1,\,\Tilde{\mathbf{x}}^2\,...,\,\Tilde{\mathbf{x}}^{n_c})\,=\,\big(\sum_{k=1}^{n_{c}}|\Tilde{\mathbf{x}}^k|\big)^{\frac{1}{2}}\,\in\, \mathbb{R}^{N \times M}, \quad \Tilde{\mathbf{x}}^k\,=\,\mathcal{F}^{-1}(\mathbf{y}^k).
\end{equation}

The problem of recovering the true image from accelerated MRI data is an inverse problem with a forward model formulated as 
\begin{equation}
    \Tilde{\mathbf{y}}^k\,=\,\mathbf{U} \circ \mathbf{y}^k\,=\,\mathbf{U} \circ \mathcal{F}(\mathbf{S}^{k} \circ \mathbf{x}) \,\,+\,\tilde{\mathbf{e}}^k, \quad k\,=\,1,\,..,\,n_c,
    \label{eq:masked_coil_kspace}
\end{equation} 
where $\tilde{\mathbf{y}}\,=\,\{\tilde{\mathbf{y}}^k\}_{k=1}^{n_c}$ denote the multi-coil accelerated measurements and $\mathbf{U}$ a subsampling mask operator (see Section \ref{par:prospect}). A solution to the inverse problem can be formulated as a variational problem
\begin{equation}
    \arg\min_{\mathbf{x}} \mathcal{D}(\tilde{\mathbf{y}}, \mathbf{x})+\mathcal{L}(\mathbf{x})
\label{eq:varprob}
\end{equation}
where $\mathcal{D}$ and $\mathcal{L}$ denote data-fidelity and regularisation terms, respectively \cite{Kaipio:1338003,mathmodelsmri}. Solving Eq. \ref{eq:varprob} is equivalent to solving the Bayesian statistical inference problem of finding the maximum a posteriori (MAP) estimation  \cite{4407762, putzky2017recurrent, Kaipio:1338003}: 
\begin{equation}
    \mathbf{x}_{\rm MAP} \,=\,\arg\max_{\mathbf{x}}\pi_{post}(\mathbf{x}|\tilde{\mathbf{y}})\,=\,\arg\max_{\mathbf{x}}\pi_{l}(\Tilde{\mathbf{y}}\,|\,\mathbf{x})\,\cdot \,\pi_{p}(\mathbf{x})\,=\,\arg \min_{\mathbf{x}} -\,\log \pi_{l}(\Tilde{\mathbf{y}}\,|\,\mathbf{x})\,-\,\log \pi_{p}(\mathbf{x}),
    \label{eq:MAP}
\end{equation}
where $\pi_{post}$ denotes the posterior distribution and $\pi_{l}$ and $\pi_{p}$ the $k$-space likelihood and prior models, respectively. Let ${\mathbf{S}^k}^*$ denote the complex conjugate of of $\mathbf{S}^k$. Assuming that $\{\tilde{\mathbf{e}}^k\}_{k=1}^{n_c}$ are i.i.d. and derived from a normal distribution  with variance $\sigma^2$, the negative log-likelihood and its gradient \cite{putzky2017recurrent,LONNING201964} are given by
\begin{equation}
    -\,\log \pi_{l}(\Tilde{\mathbf{y}}\,|\,\mathbf{x})\,=\,\frac{1}{\sigma^2}\sum_{k=1}^{n_c}\big|\big|\mathbf{U} \circ \mathcal{F}(\mathbf{S}^k \circ \mathbf{x}) -\tilde{\mathbf{y}}^k\big|\big|_2^2,
    \label{eq:likelihood}
\end{equation}
and
\begin{equation}
 \nabla_{\tilde{\mathbf{y}} | \mathbf{x}}\,:=\, \nabla_\mathbf{x}\big(-\,\log \pi_{l}(\Tilde{\mathbf{y}}\,|\,\mathbf{x})\big) \,=\,\frac{1}{\sigma^2}\sum_{k=1}^{n_c}{\mathbf{S}^k}^* \circ \mathcal{F}^{-1}\bigg(\mathbf{U} \circ \big( \mathbf{U} \circ \mathcal{F}(\mathbf{S}^k \circ \mathbf{x}) -\tilde{\mathbf{y}}^k\big)\bigg),
    \label{eq:grad_likelihood}
\end{equation}
respectively.
%%%%%%%%%%%%%%%%%%%%%%%%%%%%%%%%%%%%%%%%%%%%%%%%%%%%%%%%%%%%%%
\subsection{\textit{\lowercase{k}}-space Sampling - Subsampling}
\label{subsec:ch2sub2}
\paragraph{Prospective (Sub)sampling}
\label{par:prospect}

% \begin{figure}[h!]
% \centering
% \begin{tabular}{cccc}
% \includegraphics[width=0.2\textwidth]{figures/cartesian_dots_traj.png} &
% \includegraphics[width=0.2\textwidth]{figures/radial_dots_traj.png} &
% \includegraphics[width=0.2\textwidth]{figures/spiral_dots_traj.png} \\
% \textbf{(a)} Cartesian (Rectilinear)  & \textbf{(b)} Radial & \textbf{(c)} Spiral  \\[6pt]
% \end{tabular}
% \begin{tabular}{cccc}
% \includegraphics[width=0.2\textwidth]{figures/subsample_cartesian_dots_traj_red.png} &
% \includegraphics[width=0.2\textwidth]{figures/subsample_radial_dots_traj_red.png} \\
% \textbf{(d)} Radial Subsampling  & \textbf{(e)} Spiral Subsampling  \\[6pt]
% \end{tabular}
% \caption{$k$-space trajectories used in clinical settings.}
% \label{fig:trajectories}
% \end{figure}

\begin{figure}[ht!]
     \centering
      \begin{subfigure}[b]{0.3\textwidth}
         \centering
         \includegraphics[width=0.7\linewidth]{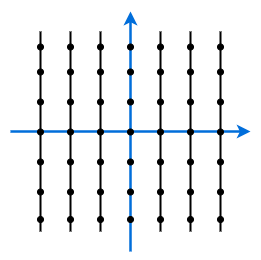}
         \caption{Cartesian (Rectilinear)}
         \label{fig:cartesian_traj}
     \end{subfigure}
     \hfill
     \begin{subfigure}[b]{0.3\textwidth}
         \centering
         \includegraphics[width=0.7\linewidth]{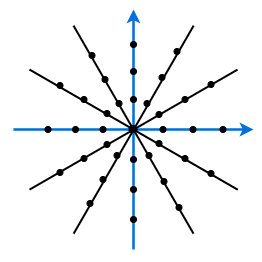}
         \caption{Radial}
         \label{fig:radial_traj}
     \end{subfigure}
     \hfill
      \begin{subfigure}[b]{0.3\textwidth}
         \centering
         \includegraphics[width=0.7\linewidth]{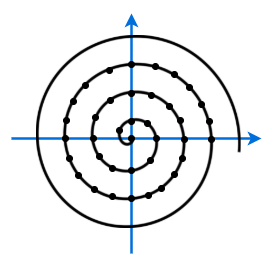}
         \caption{Spiral}
         \label{fig:spiral_traj}
     \end{subfigure}
     \medskip
     \begin{subfigure}[b]{0.3\textwidth}
         \centering
         \includegraphics[width=0.7\linewidth]{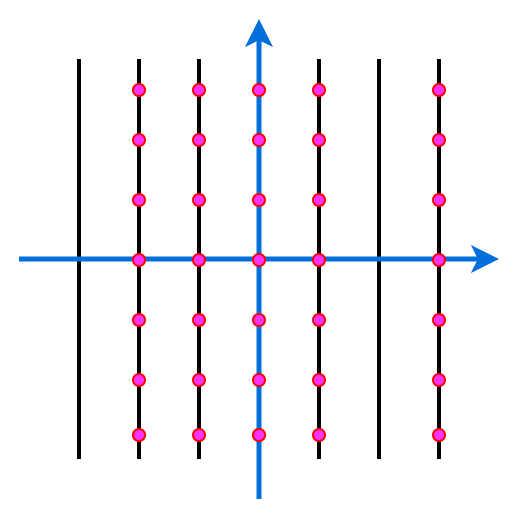}
         \caption{Cartesian Subsampling}
         \label{fig:cart_sub_traj}
     \end{subfigure}
     \begin{subfigure}[b]{0.3\textwidth}
         \centering
         \includegraphics[width=0.7\linewidth]{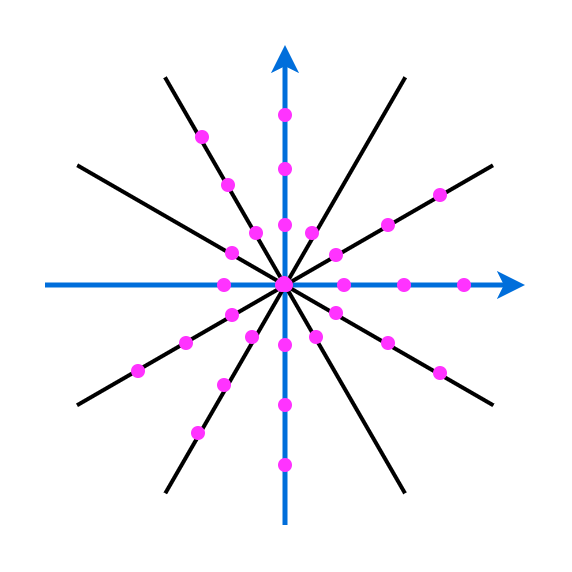}
         \caption{Radial Subsampling}
         \label{fig:radial_sub_traj}
     \end{subfigure}
     \vspace{2.0mm}
    \caption{$k$-space trajectories used in clinical settings.}
    \label{fig:trajectories}
\end{figure}

In clinical settings several sampling (and subsampling) trajectories have been used since the advent of MRI \cite{kspace_strategies, K-Spaceintheclinic}. The most conventional sampling trajectory is the Cartesian or rectilinear sampling \cite{undersampling_patterns_in_kspace}, that is, the $k$-space is measured in a line-by-line scheme on a equidistant rectangular grid (Fig. \ref{fig:cartesian_traj}). Other non-Cartesian $k$-space trajectories commonly used are the radial and spiral (Fig. \ref{fig:radial_traj} and \ref{fig:spiral_traj}). Since non-Cartesian trajectories are not acquired on an equidistant grid, one of projection reconstruction \cite{https://doi.org/10.1118/1.1677252, https://doi.org/10.1002/mrm.1910190226}, regridding \cite{PMID:18243972} or the application of the non-uniform FFT (NUFFT) is required. Cartesian subsampling is achieved by scanning some $k$-space lines and omitting others up to the chosen acceleration factor $R$ (Fig. \ref{fig:cart_sub_traj}), while Radial subsampling is achieved by scanning some $k$-space spokes and omitting others up to $R$ (Fig. \ref{fig:radial_sub_traj}).  

\paragraph{Retrospective Subsampling}
\label{par:retro}

\begin{figure}[ht!]
     \centering
      \begin{subfigure}[b]{0.4\textwidth}
         \centering
         \includegraphics[width=0.7\linewidth]{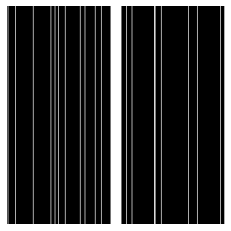}
        \caption{Rectilinear}
        \label{fig:cartesian_mask}
     \end{subfigure}
     \begin{subfigure}[b]{0.4\textwidth}
         \centering
         \includegraphics[width=0.7\linewidth]{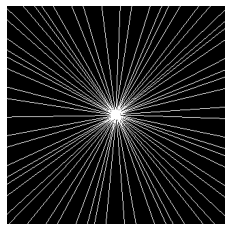}
        \caption{Radial}
        \label{fig:radial_mask}
     \end{subfigure}
\vspace{2.0mm}
\caption{Subsampling Masks on a Cartesian grid. The acceleration factor is R $\approx$ 10.}
\label{fig:masks}
\end{figure}

To simulate the subsampling performed in the MR scanner, subsampling masks were retrospectively applied onto the fully sampled $k$-space to produce a subsampled $k$-space hereby referred to as the masked or zero-filled $k$-space. More formally, let $\mathbf{U}\in\{0,1\}^{N \times M}$ denote a binary subsampling mask such that if $\mathbf{y}^k_{ij}$ is masked only if $\displaystyle \mathbf{U}_{ij}=0$, $(i,\,j)\,\in\,\{1,\,..,\, N\} \times \{1,\,..,\, M\}$. This is described by Eq. \ref{eq:masked_coil_kspace}. For each subsampling mask an acceleration factor $R$ is determined such that $\displaystyle \frac{{N\cdot M}}{{\sum_{i,j=1}^{N,M}\mathbf{U}_{ij}}}\,\approx \,R$. 

In this paper, we worked with fully sampled data acquired by Cartesian trajectories and therefore, for our experiments we generated rectilinear and radial subsampling masks on the Cartesian grid as depicted by Fig. \ref{fig:cartesian_mask} and Fig. \ref{fig:radial_mask}, respectively. 

\section{Experiments}
\label{sec:ch3}
\subsection{Implementation}
\label{subsec:ch3implementation}

\paragraph{Dataset}
\label{par:dataset}
The RIM models were trained and evaluated on the Calgary-Campinas public brain multi-coil MRI dataset which was released as part of an accelerated MRI reconstruction challenge  \cite{calgary, Beauferris2020MultiCoilMR}. The dataset is consisted of 67 three-dimensional raw $k$-space volumes that are T1-weighted, gradient-recalled echo, 1 mm isotropic sagittal and are collected on a Cartesian grid (equidistant). These amount to 10,452 slices of fully sampled $k$-spaces which we randomly split into training (40 volumes), validation (14 volumes) and test (13 volumes) sets. All data were acquired using twelve receiver coils ($n_c\,=\,12$) and the reconstructed images had $218 \times 170$ or $218 \times 180$ pixels. 

\paragraph{Subsampling}
\label{par:subsampling}
For the purposes of the experiments in this paper, two different subsampling schemes were  used to retrospectively subsample the fully acquired $k$-space data. The first scheme generated rectilinear subsampling masks by first sampling a fraction (10\% when $R=5$ or 5\% when $R=10$) of the central phase encoding columns (low frequencies) and then randomly including others up to a level of acceleration $R$ \cite{zbontar2019fastmri}. The second subsampling scheme was radial subsampling in which fully acquired Cartesian measurements  were subsampled in a radial fashion to simulate radial acquisition. Radial masks were generated using the CIRcular Cartesian UnderSampling (CIRCUS) method presented in \cite{circus} which fit our settings. Figures \ref{fig:cartesian_mask} and \ref{fig:radial_mask} illustrate an example of one of each of the subsampling masks we used. Note that the same subsampling mask was applied to all coil data and slices of a single volume sample.

\begin{figure}[ht!]
     \centering
     \includegraphics[width=1.0\linewidth]{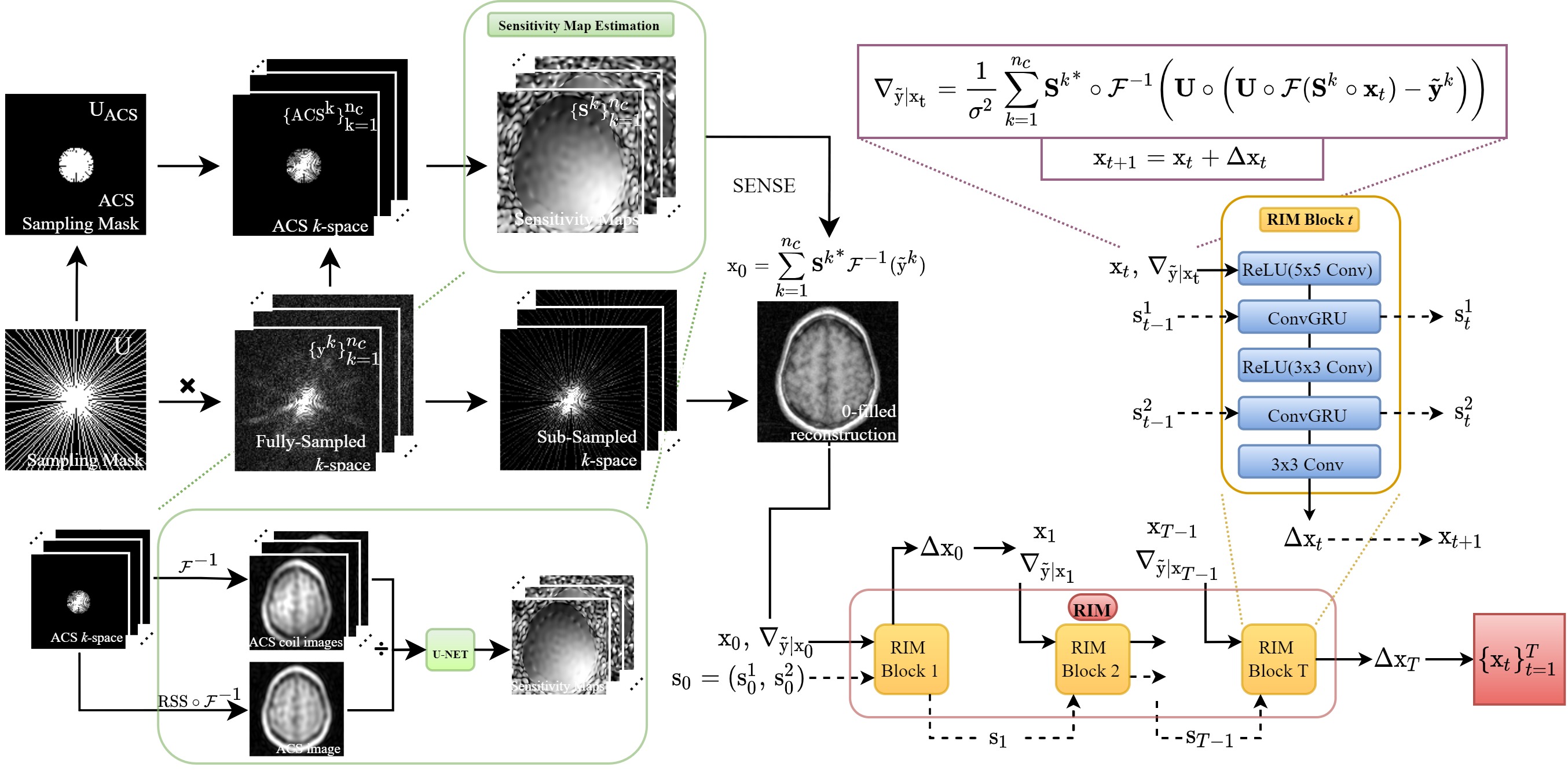}
     \vspace{2.0mm}
    \caption{RIM with radial subsampling training overview. A random radial subsampling mask is produced and applied on the multi-coil fully acquired $k$-space data. Coil sensitivity maps are produced using the ACS of the subsampled data and are refined throughout training with a U-Net. A zero-filled reconstruction using SENSE and the gradient of the negative log-likelihood are used as an input to the RIM. The internal states are used as inputs of the recurrent units. At each time-step an incremental image update is produced. The model outputs a sequence $\{\mathbf{x}_t\}_{t=1}^T$ of estimates of the true image.}
    \label{fig:rim}
\end{figure}

\paragraph{Model Architecture}
\label{par:models}
To build, train and evaluate our models, we exploited our Deep Image Reconstruction Toolkit (DIRECT) \cite{DIRECTTOOLKIT} which uses PyTorch \cite{NEURIPS2019_9015}. The architecture we opted for is the one of the Recurrent Inference Machine (RIM) as it has been shown to achieve state-of-the art performance in accelerated MRI reconstruction tasks \cite{putzky2017recurrent, putzky2019irim, LONNING201964, Beauferris2020MultiCoilMR}. At each training iteration RIMs are trained to resemble a gradient descent scheme to find the MAP estimation (Eq. \ref{eq:MAP}) using Eq. \ref{eq:likelihood} by following an iterative scheme of $T$ time-steps:
\begin{equation}
    \mathbf{s}_0=(\mathbf{0},\,\mathbf{0}),\quad \mathbf{x}_0=\sum_{k=1}^{n_c} {\mathbf{S}^k}^{*}\mathcal{F}^{-1}(\tilde{\mathbf{y}}^k),\quad \mathbf{s}_{t+1},\,\Delta \mathbf{x}_{t} = f_t(\nabla_{\tilde{\mathbf{y}}|\mathbf{x}_t}, \mathbf{x}_t, \mathbf{s}_t),\quad \mathbf{x}_{t+1}=\mathbf{x}_t+\Delta \mathbf{x}_t,\quad t=0,..,T,
\end{equation}
where $\mathbf{s}_t\,=\,(\mathbf{s}_t^1,\,\mathbf{s}_t^2)$ is called the internal state. The update function $f_t$ (RIM Block $t$) of the RIM consists of a sequence of two blocks, each consisted of a convolutional layer ($5\times 5$, $3\times 3$ kernel sizes) followed by a rectified linear unit (ReLU) activation and a convolutional gated recurrent unit (convGRU) \cite{shi2015convolutional}, followed by a $3\times 3$ convolution. At time-step $t$, $f_t$ produces an incremental update $\Delta \mathbf{x}_t$ for $\mathbf{x}_{t+1}$ and the internal state for time-step ${t+1}$. 
% At each iteration the RIM outputs a sequence $\{x_t\}_{t=1}^T$. 
Figure \ref{fig:rim} provides an overview of our model training with radial subsampling. For more details on the model architecture we refer the reader to the original paper of Recurrent Inference Machines \cite{putzky2017recurrent}.

% \newpage
\paragraph{Model Training}

\begin{wrapfigure}{r}{0.4\textwidth}
    \centering
     \includegraphics[width=0.35\linewidth]{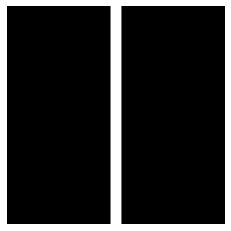}
     \includegraphics[width=0.35\linewidth]{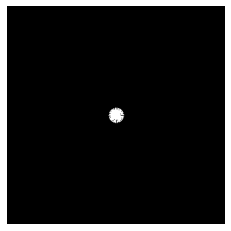}
     \caption{ACS masks.}
    \label{fig:acs_masks}
    % \vspace{-0.4cm}
\end{wrapfigure}

For our experiments we trained two RIMs. The first was trained on the radially subsampled data (Radial-RIM) and the second on the rectilinear subsampled data (Rect-RIM) employing the two subsampling schemes presented in \ref{par:subsampling} \nameref{par:subsampling}. For both RIMs we set $T=16$ time-steps and used 128 hidden channels for the convolutional layers. We also employed a U-net \cite{ronneberger2015unet} which was trained simultaneously with each RIM that sequentially improves the coil sensitivity maps. The total number of trainable parameters for each RIM amounted to 360,194 parameters and for each U-Net 484,898 parameters. Models took 2D slices of retrospectively subsampled $k$-space data as input and output 2D reconstructed images. Models were optimised using the Adam optimiser and were trained for approximately 150k iterations with a batch size of 4 slices. A warm-up schedule \cite{goyal2018accurate} was used to linearly increase the learning rate to 0.0001 over 1k warm-up iterations before decaying with a ratio of 5:1 every 5k iterations. Throughout training, the acceleration factor was arbitrarily set to 5 or 10. Every 500 iterations, the models were evaluated on the 5-fold and 10-fold accelerated validation data. Coil sensitivities were initially calculated using autocalibration data, that is, the center of the subsampled $k$-space \cite{https://doi.org/10.1002/mrm.24751}. Figure \ref{fig:acs_masks} illustrates an example of the autocalibration signal (ACS) used to calculate the sensitivity maps corresponding to the masks depicted in \ref{fig:cartesian_mask} and \ref{fig:radial_mask}. The loss function used for training the models was the sum of the $\mathcal{L}_1$ loss and a loss derived from the SSIM metric \cite{ssim}. Assuming that RSS reconstructed image $\Tilde{\mathbf{x}}$ is the ground truth image (RSS) and $\{\mathbf{x}\}_{t=1}^T$ is the model output, the loss function employed was the following
\begin{equation}
    \mathcal{R}(\mathbf{x}_T) \,=\,\frac{1}{T}\sum_{t=1}^T\big(\,\mathcal{L}_1(\Tilde{\mathbf{x}},\,\mathbf{x}_t)\,+\,\mathcal{L}_\text{SSIM}(\Tilde{\mathbf{x}},\,\mathbf{x}_t)\big)\,=\,\frac{1}{T}\sum_{t=1}^T\big(\,||\Tilde{\mathbf{x}}\,-\,\mathbf{x}_t||_1\,+\,1-\text{SSIM}(\Tilde{\mathbf{x}},\,\mathbf{x}_t)\big).
\end{equation}

\subsection{Results}
\label{subsec:ch3results}
\begin{table}[ht!]
\centering
\begin{tabular}{lclclcccl}
\toprule
\hline
 &
  \multirow{2}{*}{\textbf{Acceleration}} &
   &
  \multirow{2}{*}{\textbf{Subsampling}} &
   &
  \multicolumn{3}{c}{\textbf{Metrics}} &
   \\ \cline{6-8}
 &  &  &             &  & \textbf{PSNR $\uparrow$} & \textbf{SSIM $\uparrow$  }     & \textbf{VIF $\uparrow$  }      &  \\ \cline{2-2} \cline{4-4} \cline{6-8}
 &
  \multirow{2}{*}{5-fold} &
   & Rectilinear &  & $29.251 \pm 1.195$       & $0.867 \pm 0.018$ & $0.825 \pm 0.043$
   &
   \\
 &  &  & Radial &
   &
  \textbf{34.836 $\pm$ 0.489} &
  \textbf{0.939 $\pm$ 0.004} &
  \textbf{0.955 $\pm$ 0.006} &  \\ \cline{2-8}
 &
  \multirow{2}{*}{10-fold} &
   & Rectilinear &  & $24.969 \pm 0.755$       & $0.757 \pm 0.021$ & $0.654 \pm 0.045$
   &
   \\
 &  &  & Radial &
   &
  \textbf{31.144 $\pm$ 0.655} &
  \textbf{0.899 $\pm$ 0.009} &
  \textbf{0.905 $\pm$ 0.017} & \\ \hline
\bottomrule
\end{tabular}
\vspace{2.0mm}
\caption{Quantitative evaluation on the test data: mean metrics and standard deviations.}
\label{tab:test_metrics}
\end{table}
\noindent
To evaluate the performance of the models and the quality of the reconstructions, as well as to compare the two distinct subsampling schemes, we used three common metrics; the peak signal-to-noise ratio (PSNR), the structural similarity index meseaure (SSIM) and the visual information fidelity (VIF) \cite{ssim, zbontar2019fastmri, calgary, 1576816, 5596999}. Note that in inference time, only the last time-step $\mathbf{x}_T$ was used as the model prediction. We ran inference on the test data which were retrospectively 5-fold or 10-fold subsampled. For each acceleration factor and for each type of subsampling we selected to present results of the best model checkpoint based on the validation performance on the SSIM metric. 

\begin{figure}[ht!]
     \centering
     \includegraphics[width=1.0\linewidth]{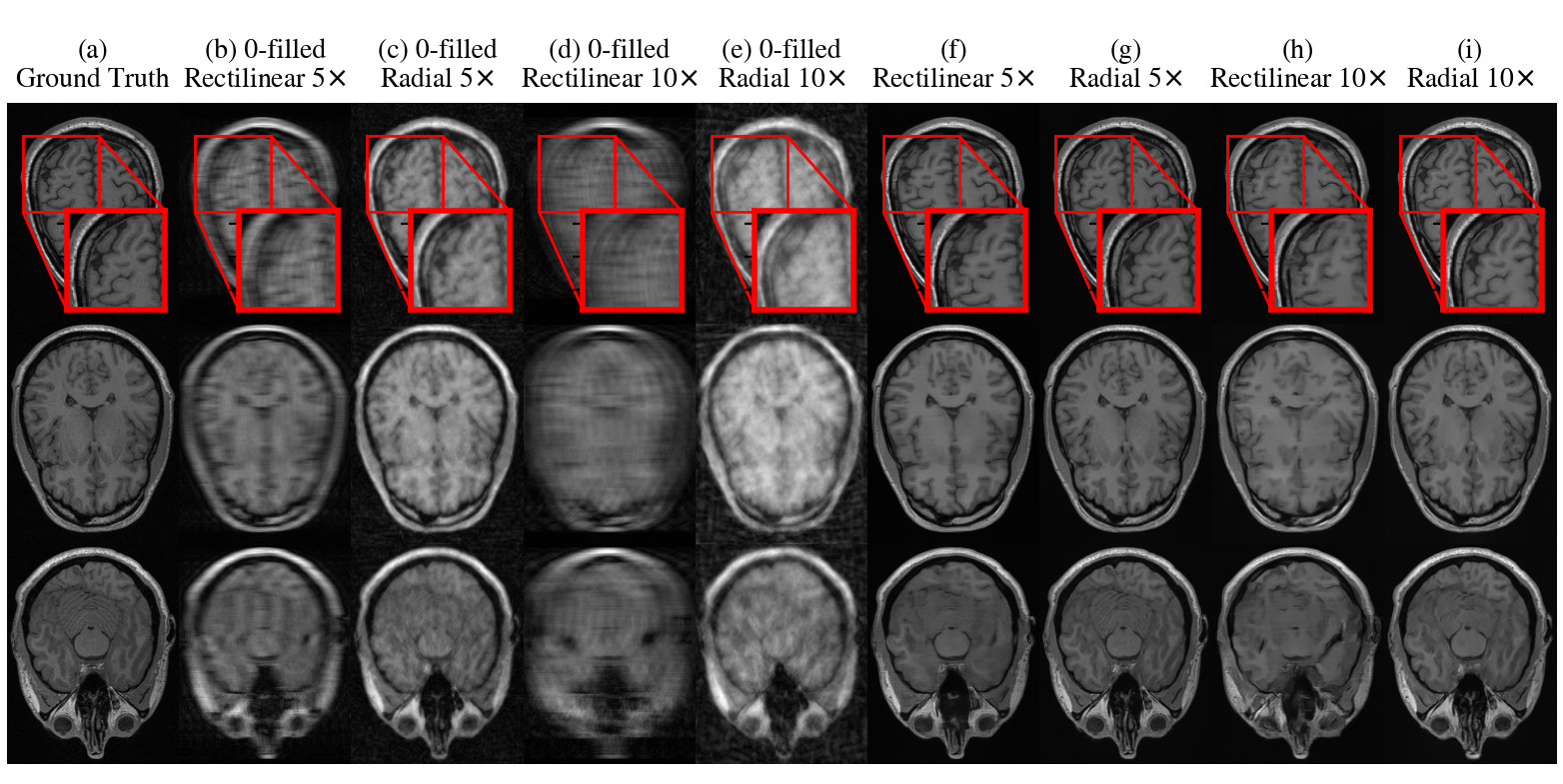}
    \caption{Visualisation of three axial slices of a brain sample from the test data. (a) Ground truth image. (b)-(e) Zero-filled RSS reconstructions. (f), (h) Rect-RIM reconstructions of 5$\times$ and 10$\times$ rectilinear accelerated data. (g), (i) Radial-RIM reconstructions of 5$\times$ and 10$\times$ radially accelerated data.}
    \label{fig:reconstructed}
\end{figure}

\begin{figure}[ht!]
     \centering
     \includegraphics[width=1.0\linewidth]{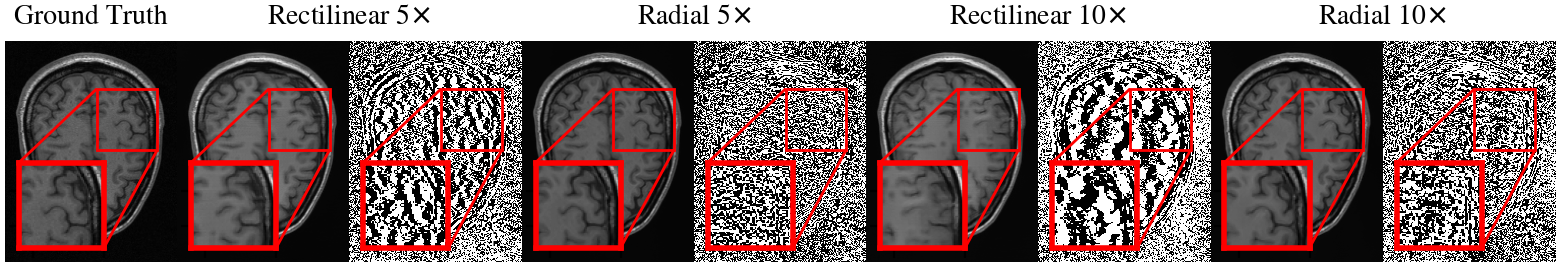}
     \vspace{0.5mm}
    \caption{Visualisation of reconstructions of a slice of a brain sample from the test data along with error maps against the ground truth image.}
    \label{fig:reconstructed_error_maps} 
\end{figure}

The quantitative results are summarised in Table \ref{tab:test_metrics}. Specifically, we report the mean metrics along with their respective standard deviations. Some qualitative results are shown by Fig. \ref{fig:reconstructed} and Fig. \ref{fig:reconstructed_error_maps} which illustrate exemplary reconstructions of axial images of brain samples from the test data. In the former are visualised image reconstructions outputted from the models, against the ground truth images and zero-filled reconstructions. The latter depicts 5$\times$, 10$\times$ accelerated image reconstructions along with error maps from the ground truth images.

As shown by Table \ref{tab:test_metrics}, Radial-RIM consistently outperformed Rect-RIM for both acceleration factors, as well as, the standard deviations of the evaluation metrics of the Radial-RIM were lower. The superiority in performance of Radial-RIM is also verified by the qualitative results as its predictions better resemble the ground truth images as they are more faithful with significantly less artifacts compared to Rect-RIM's. The zoomed regions in Figures \ref{fig:reconstructed} and \ref{fig:reconstructed_error_maps} show that it's almost impossible to distinguish which is the true image between the ground truth images and Radial-RIM's reconstructions. On the other hand, Rect-RIM's reconstructions of the 10-fold accelerated data exemplified notable artifacts which in fact, might be interpreted as pathology. 

Examining closely the error maps in Figure \ref{fig:reconstructed_error_maps}, Radial-RIM's prediction error maps resemble random noise while in Rect-RIM's prediction error maps the anatomical structure can be easily inferred as a result of the lower quality reconstructions. Another noteworthy remark of this experiment is the fact that Radial-RIM managed to infer slightly better reconstructions from the 10-fold radially subsampled data than the Rect-RIM did from 5-fold rectilinear subsampled data. 

\section{Conclusion - Discussion}
\label{sec:ch4}

In this work we have established that training Deep Learning models such as Recurrent Inference Machines on retrospectively radially subsampled data can outperform the traditional approach of rectilinear subsampling. The superior performance of radial subsampling could be attributed to the fact that radial masks allow for more incoherent sampling and sample denser when closer to the $k$-space center. The data we have used in this study were fully acquired by a conventional Cartesian trajectory, therefore, measurements were placed on an equidistant grid of $N\times M$ pixel size. In our experiments we retrospectively subsampled the data by exploiting a radial subsampling scheme as a way of simulating radial (sub)sampling in real clinical settings. Even though differences exist between the view orderings of the retrospectively and prospectively radially subsampled data, their influence requires further investigation. Additionally, in practice prospective radial (sub)sampling patterns take measurements on a non-Cartesian grid, as depicted by Fig. \ref{fig:radial_traj} and Fig. \ref{fig:radial_sub_traj}. As explained in Section \ref{sec:ch2}, this process would require regridding or the application of NUFFT. That is why our results pave the way for incorporating radial subsampling in clinical settings, which combined with Deep Learning can speed up MRI acquisitions and even allow for dynamic treatment.

\bibliography{references}
\bibliographystyle{unsrt}

\end{document}